\pdfoutput=1
\documentclass[12pt,a4paper]{article}
\usepackage{ifthen} % for conditional statements
\newboolean{pdflatex}
\setboolean{pdflatex}{true} % False for eps figures

\newboolean{articletitles}
\setboolean{articletitles}{true} % False removes titles in references

\newboolean{uprightparticles}
\setboolean{uprightparticles}{false} %True for upright particle symbols

\newboolean{inbibliography}
\setboolean{inbibliography}{false} %True once you enter the bibliography

\newboolean{forarxiv}
\setboolean{forarxiv}{true}

\usepackage{microtype}
\usepackage{lineno}  % for line numbering during review
\usepackage{xspace} % To avoid problems with missing or double spaces after
\usepackage{caption} %these three command get the figure and table captions automatically small

\usepackage{graphicx}  % to include figures (can also use other packages)
\usepackage{color}
\usepackage{colortbl}
\graphicspath{{./figs/}
              {./figs/intro/}
              {./figs/phik/}
              {./figs/kpipi/}} % Make Latex search fig subdir for figures

\usepackage{amsmath} % Adds a large collection of math symbols
\usepackage{amssymb}
\usepackage{amsbsy}
\usepackage{bm}
\usepackage{amsfonts}
\usepackage{upgreek} % Adds in support for greek letters in roman typeset

\newcommand*\patchAmsMathEnvironmentForLineno[1]{%
\expandafter\let\csname old#1\expandafter\endcsname\csname #1\endcsname
\expandafter\let\csname oldend#1\expandafter\endcsname\csname
end#1\endcsname
 \renewenvironment{#1}%
   {\linenomath\csname old#1\endcsname}%
   {\csname oldend#1\endcsname\endlinenomath}%
}
\newcommand*\patchBothAmsMathEnvironmentsForLineno[1]{%
  \patchAmsMathEnvironmentForLineno{#1}%
  \patchAmsMathEnvironmentForLineno{#1*}%
}
\AtBeginDocument{%
\patchBothAmsMathEnvironmentsForLineno{equation}%
\patchBothAmsMathEnvironmentsForLineno{align}%
\patchBothAmsMathEnvironmentsForLineno{flalign}%
\patchBothAmsMathEnvironmentsForLineno{alignat}%
\patchBothAmsMathEnvironmentsForLineno{gather}%
\patchBothAmsMathEnvironmentsForLineno{multline}%
}

\usepackage{hyperref}    % Hyperlinks in references
\usepackage[all]{hypcap} % Internal hyperlinks to floats.

\def\lhcb {\mbox{LHCb}\xspace}

\def\rich   {RICH\xspace}

\ifthenelse{\boolean{uprightparticles}}%
{

 \def\Pmu         {\ensuremath{\upmu}\xspace}

 \def\Ppi         {\ensuremath{\uppi}\xspace}

 \def\Ppsi        {\ensuremath{\uppsi}\xspace}

 \def\PDelta      {\ensuremath{\Delta}\xspace}
 \def\PXi      {\ensuremath{\Xi}\xspace}
 \def\PLambda      {\ensuremath{\Lambda}\xspace}
 \def\PSigma      {\ensuremath{\Sigma}\xspace}
 \def\POmega      {\ensuremath{\Omega}\xspace}
 \def\PUpsilon      {\ensuremath{\Upsilon}\xspace}

 \def\PB      {\ensuremath{\mathrm{B}}\xspace}
 
 \def\PD      {\ensuremath{\mathrm{D}}\xspace}

 \def\PJ      {\ensuremath{\mathrm{J}}\xspace}
 \def\PK      {\ensuremath{\mathrm{K}}\xspace}

 \def\Pb      {\ensuremath{\mathrm{b}}\xspace}
 \def\Pc      {\ensuremath{\mathrm{c}}\xspace}

 \def\Pi      {\ensuremath{\mathrm{i}}\xspace}

 \def\Pp      {\ensuremath{\mathrm{p}}\xspace}

 \def\Ps      {\ensuremath{\mathrm{s}}\xspace}

}
{

 \def\Pmu         {\ensuremath{\mu}\xspace}

 \def\Ppi         {\ensuremath{\pi}\xspace}

 \def\Ppsi        {\ensuremath{\psi}\xspace}
 
 \mathchardef\PDelta="7101
 \mathchardef\PXi="7104
 \mathchardef\PLambda="7103
 \mathchardef\PSigma="7106
 \mathchardef\POmega="710A
 \mathchardef\PUpsilon="7107
 
 \def\PB      {\ensuremath{B}\xspace}
 
 \def\PD      {\ensuremath{D}\xspace}

 \def\PJ      {\ensuremath{J}\xspace}
 \def\PK      {\ensuremath{K}\xspace}

 \def\Pb      {\ensuremath{b}\xspace}
 \def\Pc      {\ensuremath{c}\xspace}

 \def\Pi      {\ensuremath{i}\xspace}

 \def\Pp      {\ensuremath{p}\xspace}

 \def\Ps      {\ensuremath{s}\xspace}

}

\makeatletter
\ifcase \@ptsize \relax% 10pt
  \newcommand{\miniscule}{\@setfontsize\miniscule{4}{5}}% \tiny: 5/6
\or% 11pt
  \newcommand{\miniscule}{\@setfontsize\miniscule{5}{6}}% \tiny: 6/7
\or% 12pt
  \newcommand{\miniscule}{\@setfontsize\miniscule{5}{6}}% \tiny: 6/7
\fi
\makeatother

\DeclareRobustCommand{\optbar}[1]{\shortstack{{\miniscule (\rule[.5ex]{1.25em}{.18mm})}
  \\ [-.7ex] $#1$}}

\def\mup        {{\ensuremath{\Pmu^+}}\xspace}
\def\mun        {{\ensuremath{\Pmu^-}}\xspace} % muon negative (\mum is taken)
\def\mumu       {{\ensuremath{\Pmu^+\Pmu^-}}\xspace}

\def\squark    {{\ensuremath{\Ps}}\xspace}

\def\cquark    {{\ensuremath{\Pc}}\xspace}

\def\bquark    {{\ensuremath{\Pb}}\xspace}

\def\pion   {{\ensuremath{\Ppi}}\xspace}

\def\pip    {{\ensuremath{\pion^+}}\xspace}
\def\pim    {{\ensuremath{\pion^-}}\xspace}

\def\kaon    {{\ensuremath{\PK}}\xspace}
  \def\Kbar    {{\kern 0.2em\overline{\kern -0.2em \PK}{}}\xspace}

\def\KorKbar    {\kern 0.18em\optbar{\kern -0.18em K}{}\xspace}

\def\Kp      {{\ensuremath{\kaon^+}}\xspace}
\def\Km      {{\ensuremath{\kaon^-}}\xspace}

\def\Kstarz  {{\ensuremath{\kaon^{*0}}}\xspace}

  \def\Dbar    {{\kern 0.2em\overline{\kern -0.2em \PD}{}}\xspace}
\def\D       {{\ensuremath{\PD}}\xspace}

\def\DorDbar    {\kern 0.18em\optbar{\kern -0.18em D}{}\xspace}
\def\Dz      {{\ensuremath{\D^0}}\xspace}
\def\Dzb     {{\ensuremath{\Dbar{}^0}}\xspace}

\def\Dstarp  {{\ensuremath{\D^{*+}}}\xspace}

\def\B       {{\ensuremath{\PB}}\xspace}
\def\Bbar    {{\ensuremath{\kern 0.18em\overline{\kern -0.18em \PB}{}}}\xspace}

\def\BorBbar    {\kern 0.18em\optbar{\kern -0.18em B}{}\xspace}
\def\Bz      {{\ensuremath{\B^0}}\xspace}

\def\Bu      {{\ensuremath{\B^+}}\xspace}

\def\Bp      {{\ensuremath{\Bu}}\xspace}

\def\Bd      {{\ensuremath{\B^0}}\xspace}

\def\jpsi     {{\ensuremath{{\PJ\mskip -3mu/\mskip -2mu\Ppsi\mskip 2mu}}}\xspace}
\def\psitwos  {{\ensuremath{\Ppsi{(2S)}}}\xspace}

  \def\Y#1S{\ensuremath{\PUpsilon{(#1S)}}\xspace}% no space before {...}!

\def\proton      {{\ensuremath{\Pp}}\xspace}

\def\Lbar        {{\ensuremath{\kern 0.1em\overline{\kern -0.1em\PLambda}}}\xspace}
\def\LorLbar    {\kern 0.18em\optbar{\kern -0.18em \PLambda}{}\xspace}

\def\BF         {{\ensuremath{\cal B}}\xspace}

\newcommand{\decay}[2]{\ensuremath{#1\!\to #2}\xspace}         % {\Pa}{\Pb \Pc}

\def\to                 {\ensuremath{\rightarrow}\xspace}

\def\qsq       {{\ensuremath{q^2}}\xspace}

\def\AT#1     {\ensuremath{A_{\mathrm{T}}^{#1}}\xspace}           % 2

\def\C#1      {\ensuremath{\mathcal{C}_{#1}}\xspace}                       % 9
\def\Cp#1     {\ensuremath{\mathcal{C}_{#1}^{'}}\xspace}                    % 7
\def\Ceff#1   {\ensuremath{\mathcal{C}_{#1}^{\mathrm{(eff)}}}\xspace}        % 9
\def\Cpeff#1  {\ensuremath{\mathcal{C}_{#1}^{'\mathrm{(eff)}}}\xspace}       % 7
\def\Ope#1    {\ensuremath{\mathcal{O}_{#1}}\xspace}                       % 2
\def\Opep#1   {\ensuremath{\mathcal{O}_{#1}^{'}}\xspace}                    % 7

\newcommand{\tev}{\ifthenelse{\boolean{inbibliography}}{\ensuremath{~T\kern -0.05em eV}\xspace}{\ensuremath{\mathrm{\,Te\kern -0.1em V}}}\xspace}
\newcommand{\gev}{\ensuremath{\mathrm{\,Ge\kern -0.1em V}}\xspace}
\newcommand{\mev}{\ensuremath{\mathrm{\,Me\kern -0.1em V}}\xspace}
\newcommand{\kev}{\ensuremath{\mathrm{\,ke\kern -0.1em V}}\xspace}
\newcommand{\ev}{\ensuremath{\mathrm{\,e\kern -0.1em V}}\xspace}
\newcommand{\gevc}{\ensuremath{{\mathrm{\,Ge\kern -0.1em V\!/}c}}\xspace}
\newcommand{\mevc}{\ensuremath{{\mathrm{\,Me\kern -0.1em V\!/}c}}\xspace}
\newcommand{\gevcc}{\ensuremath{{\mathrm{\,Ge\kern -0.1em V\!/}c^2}}\xspace}
\newcommand{\gevgevcccc}{\ensuremath{{\mathrm{\,Ge\kern -0.1em V^2\!/}c^4}}\xspace}
\newcommand{\mevcc}{\ensuremath{{\mathrm{\,Me\kern -0.1em V\!/}c^2}}\xspace}

\def\mum  {\ensuremath{{\,\upmu\rm m}}\xspace}

\def\invfb   {\ensuremath{\mbox{\,fb}^{-1}}\xspace}

\def\gsim{{~\raise.15em\hbox{$>$}\kern-.85em
          \lower.35em\hbox{$\sim$}~}\xspace}
\def\lsim{{~\raise.15em\hbox{$<$}\kern-.85em
          \lower.35em\hbox{$\sim$}~}\xspace}

\def\ptot       {\mbox{$p$}\xspace}
\def\pt         {\mbox{$p_{\rm T}$}\xspace}

\def\evtgen     {\mbox{\textsc{EvtGen}}\xspace}

\def\geant      {\mbox{\textsc{Geant4}}\xspace}

\def\photos     {\mbox{\textsc{Photos}}\xspace}

\def\pythia     {\mbox{\textsc{Pythia}}\xspace}

\def\tell1  {TELL1\xspace}
\def\ukl1   {UKL1\xspace}

\newcommand{\comment}[1]{}

\def\pz{\phantom{0}}
\def\pzz{\phantom{00}}

\def\Kone{\ensuremath{K_{1}}\xspace}
\def\Koneb{{\kern 0.2em\overline{\kern -0.2em K}{}}_{1}\xspace} % no ensuremath

\def\thetakone{\ensuremath{\theta_{\Kone}}\xspace}

\def\pipi{\ensuremath{\pip\pim}\xspace}

\def\kpipi{{\ensuremath{\Kp\pip\pim}}\xspace}

\def\btokpipimumu{\ensuremath{\mbox{\decay{\Bp}{\Kp\pip\pim\mumu}}}\xspace}
\def\btokpipijpsi{\ensuremath{\mbox{\decay{\Bp}{\jpsi\Kp\pip\pim}}}\xspace}

\def\btojpsikpipi{\ensuremath{\mbox{\decay{\Bp}{\jpsi\Kp\pip\pim}}}\xspace}
\def\btopsitwosk{\ensuremath{\mbox{\decay{\Bp}{\psitwos\Kp}}}\xspace}
\def\btophikmumu{\ensuremath{\decay{\Bp}{\phi\Kp\mumu}}\xspace}
\def\btophikjpsi{\ensuremath{\decay{\Bp}{\jpsi\phi\Kp}}\xspace}

\def\btojpsiphik{\ensuremath{\decay{\Bp}{\jpsi\phi\Kp}}\xspace}

\def\btokonemumu{\ensuremath{\mbox{\decay{\Bp}{K_1(1270)^+\mumu}}}\xspace}

\def\Tm  {\ensuremath{{\,\mathrm{Tm}}}\xspace}

\def\dBF{\ensuremath{{\rm d}\BF}\xspace}
\def\dqsq{\ensuremath{{\rm d}\qsq}\xspace}

\def\sig{\ensuremath{\mathrm{sig}}\xspace}
\def\norm{\ensuremath{\mathrm{norm}}\xspace}

\newcommand{\pergevgevcccc}{\ensuremath{{\mathrm{\,Ge\kern -0.1em V^{-2}}c^{4}}}\xspace}

\newcommand{\e}[1]{\ensuremath{\times 10^{#1}}}

\usepackage{cite} % Allows for ranges in citations
\usepackage{mciteplus}

\usepackage{longtable} % only for template; not usually to be used in PAPERs

\begin{document}

%%%%%%%%%%%%%%%%%%%%%%%%%
%%%%% Title     %%%%%%%%%
%%%%%%%%%%%%%%%%%%%%%%%%%
\renewcommand{\thefootnote}{\fnsymbol{footnote}}
\setcounter{footnote}{1}

% %%%%%%% CHOOSE TITLE PAGE--------
% $Id: title-LHCb-PAPER.tex 61978 2014-10-15 11:02:32Z shall $
% ===============================================================================
% Purpose: LHCb-PAPER journal paper title page template
% Author:
% Created on: 2010-09-25
% ===============================================================================

%%%%%%%%%%%%%%%%%%%%%%%%%
%%%%%  TITLE PAGE  %%%%%%
%%%%%%%%%%%%%%%%%%%%%%%%%
\begin{titlepage}
\pagenumbering{roman}

% Header ---------------------------------------------------
\vspace*{-1.5cm}
\centerline{\large EUROPEAN ORGANIZATION FOR NUCLEAR RESEARCH (CERN)}
\vspace*{1.5cm}
\hspace*{-0.5cm}
\begin{tabular*}{\linewidth}{lc@{\extracolsep{\fill}}r}
\ifthenelse{\boolean{pdflatex}}% Logo format choice
{\vspace*{-2.7cm}\mbox{\!\!\!\includegraphics[width=.14\textwidth]{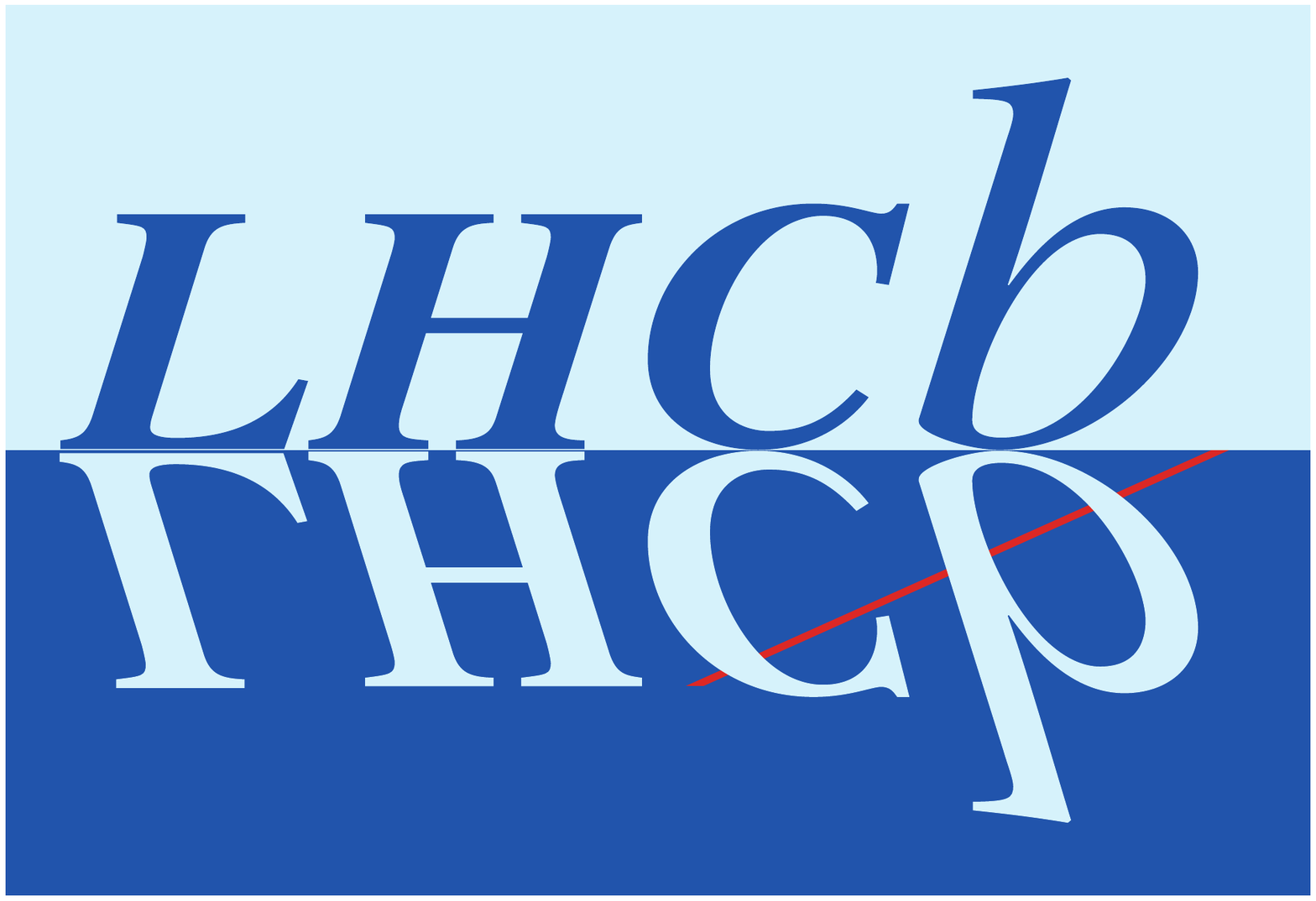}} & &}%
{\vspace*{-1.2cm}\mbox{\!\!\!\includegraphics[width=.12\textwidth]{lhcb-logo.eps}} & &}%
\\
 & & CERN-PH-EP-2014-193 \\  % ID
 & & LHCb-PAPER-2014-030 \\  % ID
 %& & \today \\ % Date - Can also hardwire e.g.: 23 March 2010
 & & 5 August 2014 \\ % Date - Can also hardwire e.g.: 23 March 2010
 & & \\
% not in paper \hline
\end{tabular*}

%\vspace*{4.0cm}
\vspace*{2.2cm}

% Title --------------------------------------------------
{\bf\boldmath\huge
\begin{center}
  First observations of the rare decays
  $B^+\!\rightarrow K^+\pi^+\pi^-\mu^+\mu^-$ and
  $B^+\!\rightarrow\phi K^+\mu^+\mu^-$
\end{center}
}

\vspace*{1.0cm}

% Authors -------------------------------------------------
\begin{center}
The LHCb collaboration\footnote{Authors are listed at the end of this paper.}
\end{center}

\vspace{\fill}

% Abstract -----------------------------------------------
\begin{abstract}
  \noindent
  First observations of the rare decays
\mbox{$B^+\!\rightarrow K^+\pi^+\pi^-\mu^+\mu^-$} and
\mbox{$B^+\!\rightarrow \phi K^+\mu^+\mu^-$} are presented using data corresponding to an integrated luminosity of $3.0\invfb$, collected
by the LHCb experiment at centre-of-mass energies of
$7$ and $8\tev$.
The branching fractions of the decays are
\begin{align*}
  \mathcal{B}(B^+\!\!\rightarrow\!K^+\pi^+\pi^-\mu^+\mu^-) &=
  \left(4.36\,^{+0.29}_{-0.27}\,\mathrm{(stat)}\pm 0.21\,\mathrm{(syst)\pm0.18\,\mathrm{(norm)}}\right)\times10^{-7},\\
  \mathcal{B}(B^+\!\!\rightarrow\!\phi K^+\mu^+\mu^-) &=
  \left(0.82 \,^{+0.19}_{-0.17}\,\mathrm{(stat)}\,^{+0.10}_{-0.04}\,\mathrm{(syst)}\pm0.27\,\mathrm{(norm)}\right)  \times10^{-7},
\end{align*}
where the uncertainties are statistical, systematic, and due to the
uncertainty on the branching fractions of the normalisation modes.
A measurement of the differential branching fraction
in bins of the invariant mass squared of the dimuon system is also presented for
the decay \mbox{$\Bp\!\to\Kp\pip\pim\mup\mun$}.

\end{abstract}

\vspace*{1.0cm}

\begin{center}
  Published in \href{http://dx.doi.org/10.1007/JHEP10(2014)064}{JHEP 10\,(2014)\,064}
\end{center}

\vspace{\fill}

{\footnotesize
\centerline{\copyright~CERN on behalf of the \lhcb collaboration, license \href{http://creativecommons.org/licenses/by/4.0/}{CC-BY-4.0}.}}
\vspace*{2mm}

\end{titlepage}

%%%%%%%%%%%%%%%%%%%%%%%%%%%%%%%%
%%%%%  EOD OF TITLE PAGE  %%%%%%
%%%%%%%%%%%%%%%%%%%%%%%%%%%%%%%%

%  empty page follows the title page ----
\newpage
\setcounter{page}{2}
\mbox{~}

% Author List ----------------------------
%  You need to get a new author list!

%\newpage
%\input{LHCb_authorlist.tex}

%The author list for journal publications is provided by the Membership Committee shortly after 'approval to go to paper' has been given.
%It will be made available on the page
%\verb!http://www.physik.uzh.ch/~strauman/forMemCo/LHCb-PAPER-XXXX-XXX/! .
%It will be sent to you by email shortly after a paper number has beens assigned.
%The author list should be included already at first circulation,
%to allow new members of the collaboration to verify whether they have been included correctly.
%Occasionally a misspelled name is corrected or associated institutions become full members.
%In that case, a new author list will be sent to you.
%In case line numbering doesn't work well after including the authorlist, try moving the \verb!\bigskip! after the last author to a separate line.

%The authorship for Conference Reports should be ``The LHCb
  %collaboration'', with a footnote giving the name(s) of the contact
  %author(s), but without the full list of collaboration names.

\cleardoublepage

% %%%%%%%%%%%%% ---------

\renewcommand{\thefootnote}{\arabic{footnote}}
\setcounter{footnote}{0}

%%%%%%%%%%%%%%%%%%%%%%%%%
%%%%% Main text %%%%%%%%%
%%%%%%%%%%%%%%%%%%%%%%%%%

\pagestyle{plain} % restore page numbers for the main text
\setcounter{page}{1}
\pagenumbering{arabic}

%% Uncomment during review phase.
%% Comment before a final submission.
%\linenumbers

\section{Introduction}
\label{sec:intro}
The \btokpipimumu and \btophikmumu decays
proceed via $\bquark\!\to\squark$ flavour changing neutral currents (FCNC).\footnote{
Charge conjugation is implied throughout this paper.}
In the Standard Model (SM), FCNC decays are forbidden
at the tree level
and are only allowed as higher-order electroweak loop processes.
In extensions of the SM, new particles can
significantly change the branching fractions and angular distributions of the observed final-state particles.
Due to their sensitivity to effects beyond the SM,
semileptonic $B$ decays involving FCNC transitions are currently under intense study at
the LHCb experiment~\cite{LHCb-PAPER-2013-019,LHCb-PAPER-2013-037,LHCb-PAPER-2013-017,LHCb-PAPER-2014-006}.

The \kpipi system in the final state of the \btokpipimumu\ decay can result from the decay of several strange resonances.
Its composition was studied by the Belle collaboration for the tree-level decay $\decay{\Bp}{\jpsi(\to\mumu)\Kp\pip\pim}$~\cite{Guler:2010if},
where the $K_1(1270)^+$ meson was found to have a prominent contribution.
The $K_1(1270)^+$ and the $K_1(1400)^+$ mesons are the mass eigenstates that result from mixing of the
$P$-wave axial vector mesons $^3P_1$ ($K_{1A}$) and $^1P_1$ ($K_{1B}$)
with the mixing angle \thetakone~\cite{PhysRevD.47.1252}.
The value of \thetakone is either about $-33^\circ$ or
$-57^\circ$~\cite{PhysRevD.47.1252,Tayduganov:2011ui,Hatanaka:2008xj,Cheng:2011pb,Divotgey:2013jba,Cheng:2013cwa}
with most recent determinations favouring the former~\cite{Hatanaka:2008xj,Cheng:2011pb,Divotgey:2013jba,Cheng:2013cwa}.
The decay \btojpsiphik was first observed by the CLEO collaboration~\cite{Jessop:1999cr} and recently investigated in the search for the
$X(4140)$~\cite{Aaltonen:2009tz, Aaij:2012pz, Abazov:2013xda, Chatrchyan:2013dma}.

The branching fraction of the rare decay \btokonemumu,
which is
expected to contribute significantly to the $\Kp\pip\pim\mup\mun$ final-state,
is predicted to be
$\BF(\btokonemumu) = (2.3\,^{+1.3}_{-1.0}\,^{+0.0}_{-0.2})\e{-6}$~\cite{Hatanaka:2008gu}.
Here, the first uncertainty originates from the form-factor calculations, while
the second is from the uncertainty on the mixing angle \thetakone.
However, due to the unknown resonance structure of the final-state hadrons,
there are no inclusive theoretical predictions available for
the branching fractions of the decays $\btokpipimumu$ and $\btophikmumu$.

This paper presents the first observations of the decays \btokpipimumu and
$\decay{\Bp}{\phi(1020)\Kp\mumu}$, using a data sample collected by the LHCb experiment, corresponding to an integrated luminosity of $3.0\invfb$.
The data were recorded in the years 2011 and 2012 at
centre-of-mass energies of 7 and 8\tev, respectively.
In addition, a measurement of the differential branching fraction
$\dBF(\btokpipimumu)/\dqsq$, where \qsq is the invariant mass
squared of the dimuon system, is presented.

\section{The LHCb detector}
\label{sec:Detector}
The \lhcb detector \cite{Alves:2008zz} is a single-arm forward spectrometer
covering the pseudorapidity range $2<\eta<5$, designed for the study of particles
containing \bquark or \cquark quarks.
The detector includes a high-precision tracking system consisting of a
silicon-strip vertex detector surrounding the $\proton\proton$ interaction
region, a large-area silicon-strip detector located upstream of a dipole magnet
with a bending power of about 4\Tm, and three stations of silicon-strip
detectors and straw drift tubes placed downstream.
The tracking system provides a measurement of momentum, \ptot,  with
a relative uncertainty that varies from 0.4\% at low momentum to 0.6\% at 100\gevc. 
The minimum distance of a track to a primary $pp$ interaction vertex (PV), the impact parameter (IP), is measured with a resolution of $(15+29/\pt)\mum$,
where \pt is the the component of \ptot transverse to the beam, in \gevc. 
Charged hadrons are identified using two ring-imaging Cherenkov detectors
(\rich) \cite{LHCb-DP-2012-003}.
Photon, electron and hadron candidates are identified by a calorimeter system
consisting of scintillating-pad and preshower detectors, an
electromagnetic calorimeter and a hadronic calorimeter. Muons are identified by
a system composed of alternating layers of iron and multiwire proportional
chambers \cite{LHCb-DP-2012-002}.
The trigger \cite{LHCb-DP-2012-004} consists of a hardware stage, based on
information from the calorimeter and muon systems, followed by a software stage,
which applies a full event reconstruction.

Simulated events 
are used to determine trigger, reconstruction and selection efficiencies. 
In addition, simulated samples are used to estimate possible backgrounds from $B$ meson decays that can mimic the final states of the signal decays. 
Simulated events are generated using 
\pythia~\cite{Sjostrand:2006za,*Sjostrand:2007gs} with a specific \lhcb
configuration~\cite{LHCb-PROC-2010-056}.
Decays of hadronic particles are described by \evtgen~\cite{Lange:2001uf}, in
which final-state radiation is generated using \photos~\cite{Golonka:2005pn}.
The interaction of the generated particles with the detector and its response
are implemented using the \geant toolkit~\cite{Allison:2006ve,*Agostinelli:2002hh}
as described in Ref.~\cite{LHCb-PROC-2011-006}.

\section{Selection of signal candidates}
\label{sec:sel}
The $\btokpipimumu$ and $\btophikmumu$ signal candidates are first required to pass the hardware trigger stage, which selects
muons with $\pt>1.76\gevc$.
In the subsequent software trigger stage, at least one of the final-state hadrons (muons) is
required to have both $\pt>1.6\gevc$ ($1.0\gevc$) and IP larger than $100\mum$ with 
respect to any PV in the event. 
A multivariate algorithm~\cite{BBDT} is used to identify secondary vertices that are consistent with the decay of a $b$ hadron with muons in the final state.

Signal candidates are formed by combining two muons of opposite charge with
three charged hadrons. 
Reconstructed signal candidate tracks must have significant displacement from any PV in the event.
The signal candidate tracks are required to form a secondary vertex of good fit quality which is significantly displaced from the PV. 
Particle identification information from the \rich\ detectors (PID) is used to 
identify the final-state hadrons. 
For \btokpipimumu\ decays, 
the invariant mass of the $\Kp\pip\pim$ system is required to be 
below $2400\mevcc$. 
For \btophikmumu\ decays with $\decay{\phi}{\Kp\Km}$, the
invariant mass of the $\Kp\Km$ system is required to be within $12\mevcc$ of the
known $\phi$ meson mass~\cite{PDG2012}.
This mass region contains almost entirely $\decay{\phi}{\Kp\Km}$ 
meson decays with negligible background. 

The final states of the signal decays can be mimicked by other $B$
decays, which represent potential sources of background.
Resonant decays, where the muon pair originates from either \jpsi or \psitwos\ meson decays, are
removed by rejecting events where the invariant mass of the dimuon system is in
the veto regions $2946<m(\mumu)<3176\mevcc$ or $3586<m(\mumu)<3766\mevcc$.
The radiative tails of the $\jpsi$ ($\psitwos$) decays are suppressed by extending
the lower edge of these veto regions 
down by $250\mevcc$ ($100\mevcc$) if the reconstructed $\Bp$ mass is
smaller than $5230\mevcc$.
In the mass region $5330<m(\Bp)<5450\mevcc$ the
upper edge of the
vetoes is extended up by
$40\mevcc$ to reject a small fraction of misreconstructed $\jpsi$ and
$\psitwos$ meson decays.
The resonant decays can also be misreconstructed as signal if a muon from the
charmonium decay is misidentified as a hadron and vice versa.
To remove this potential background the invariant mass of the $\mup\pim$ or $\mup\Km$ system
is calculated 
assigning the muon mass to the hadron. 
If the mass falls within 50\mevcc of the known \jpsi or \psitwos
masses~\cite{PDG2012}, the candidate is rejected.

Potential background from the electroweak-penguin decay $\decay{\Bz}{\Kstarz\mumu}$,
where the $\decay{\Kstarz}{\Kp\pim}$ decay is combined with a random $\pip$ meson, is studied and
found to be negligible.
Backgrounds from semileptonic $b\to c(\to s\mup\nu_\mu) \mun\bar{\nu}_\mu$ cascade decays,
as well as fully hadronic $B$ decays such as $\decay{\Bp}{\Dzb(\to\Kp\pipi\pim)\pip}$
where two hadrons are misidentified as muons, are also
negligible.

Combinatorial background is suppressed with a boosted decision tree
(BDT)~\cite{Breiman,AdaBoost}. 
The BDT training uses 
{\it sWeighted}~\cite{2005NIMPA.555..356P} candidates from the control channel \btojpsikpipi as a signal proxy
and the high $\Bu$ mass sideband ($5529 < m(\kpipi\mumu) < 5780\mevcc$) of $\btokpipimumu$ candidates as a background proxy.
The BDT uses geometric and kinematic variables in the training,
including the \pt\ of the final state tracks and their displacement from the PV.
Additionally, the \pt\ of the reconstructed $\Bu$ candidate, as well as information on the quality of the decay vertex and its displacement are used.
Requirements on the BDT response
and the PID criteria, which discriminate between kaons and pions for the reconstructed final-state hadrons, 
are optimised
simultaneously using the metric $S/\sqrt{S+B}$.
Here, $S$ and $B$ denote the expected signal and background yields.
The value of $S$ is calculated using an estimate for the branching fraction of the decay \btokonemumu. 
This branching fraction is determined by scaling 
that of the rare decay $\decay{\Bd}{\Kstarz\mup\mun}$~\cite{LHCb-PAPER-2013-019}
by the branching fraction ratio of the radiative decays $\decay{\Bp}{K_1(1270)^+\gamma}$ and $\decay{\Bd}{\Kstarz\gamma}$~\cite{PDG2012}.

To determine the branching fractions of the signal decays, the normalisation modes \btopsitwosk, with the subsequent decay
$\decay{\psitwos}{\jpsi(\to\mumu)\pip\pim}$,
and \btophikjpsi are used.
The branching fraction of the decay $\btopsitwosk$ is $(6.27 \pm 0.24)\e{-4}$~\cite{PDG2012}, 
and the branching fraction of the decay $\btojpsiphik$ is $(5.2 \pm 1.7)\e{-5}$~\cite{PDG2012}.
The final states of the normalisation modes are identical to those of the signal decays, which is beneficial since many systematic effects are expected to cancel.
Both normalisation modes are selected in analogy to the signal
decays
except for additional mass requirements.
For the \psitwos decay, the reconstructed $\pipi\mumu$ mass is required to be within $60\mevcc$ of the known \psitwos\ mass. 
The reconstructed invariant mass of the dimuon system originating from the $\jpsi$ meson decay is required to be within $50\mevcc$ of the known $\jpsi$ mass.

\section[Differential branching fraction of the decay \btokpipimumu]
{Differential branching fraction of the decay
  \mbox{$\boldsymbol{B^+\!\rightarrow K^+\pi^+\pi^-\mu^+\mu^-}$}}
\label{sec:kpipi}
\begin{figure}
  \begin{center}
    \includegraphics[width=0.49\textwidth]{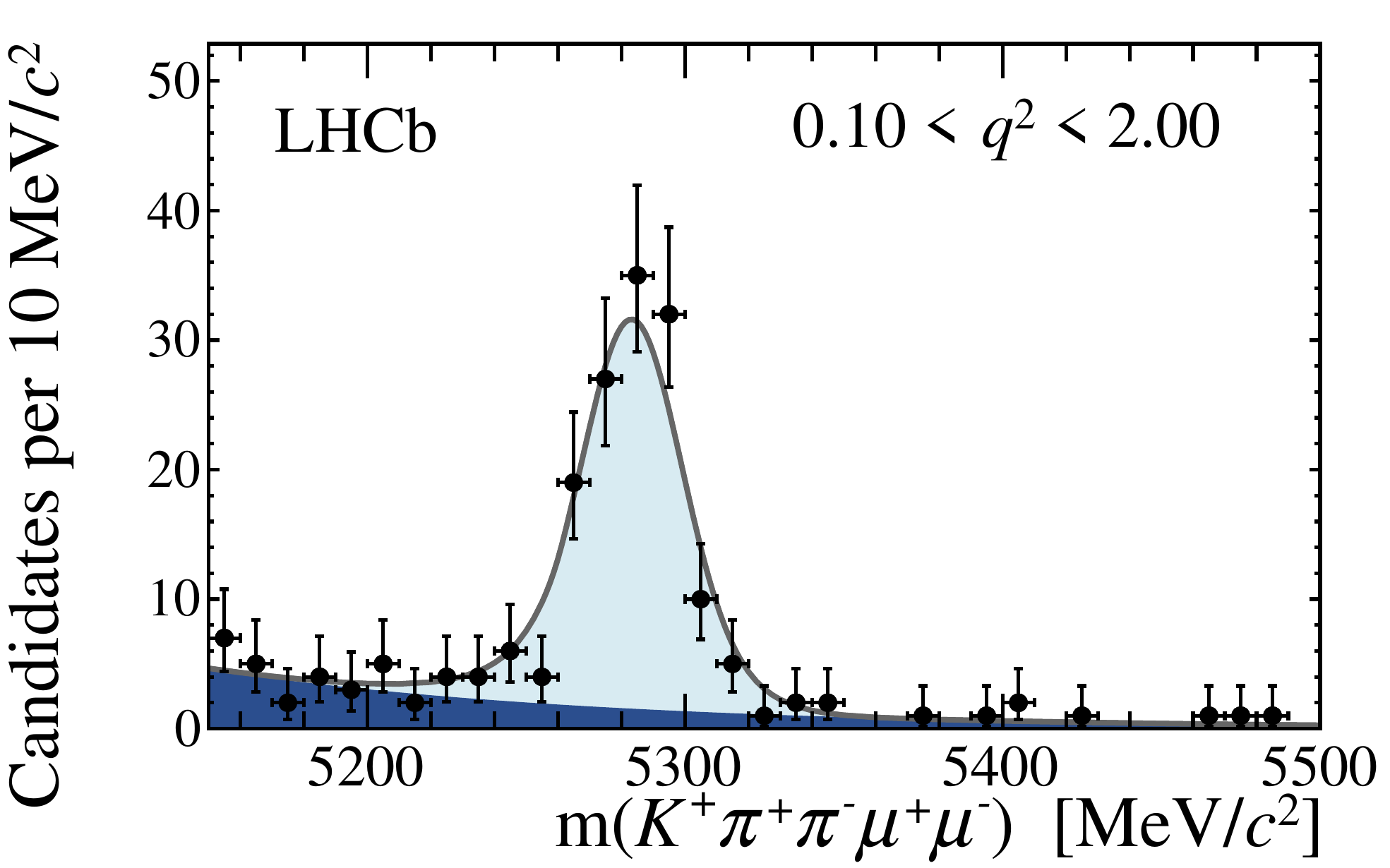}
    \includegraphics[width=0.49\textwidth]{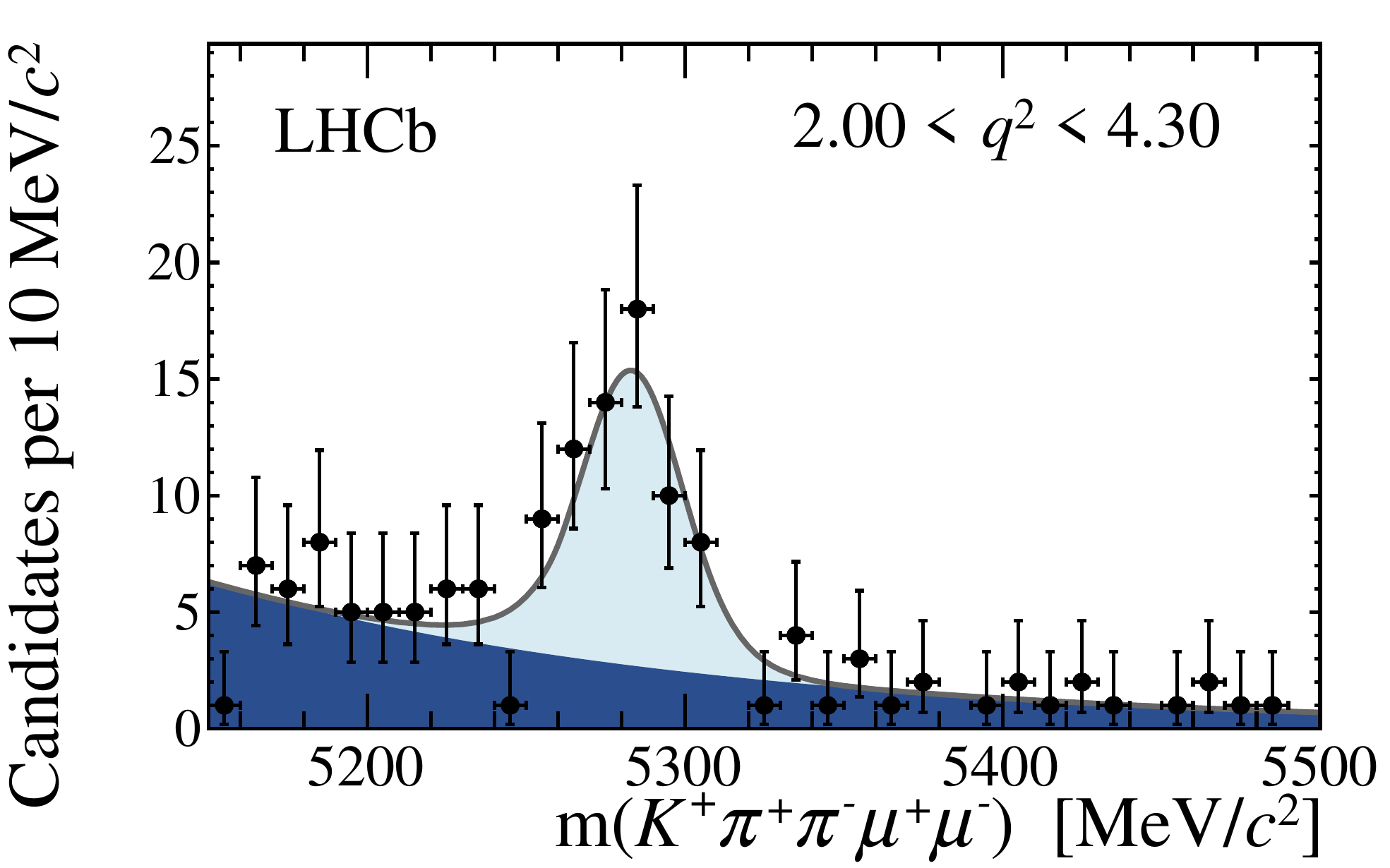}\\
    \includegraphics[width=0.49\textwidth]{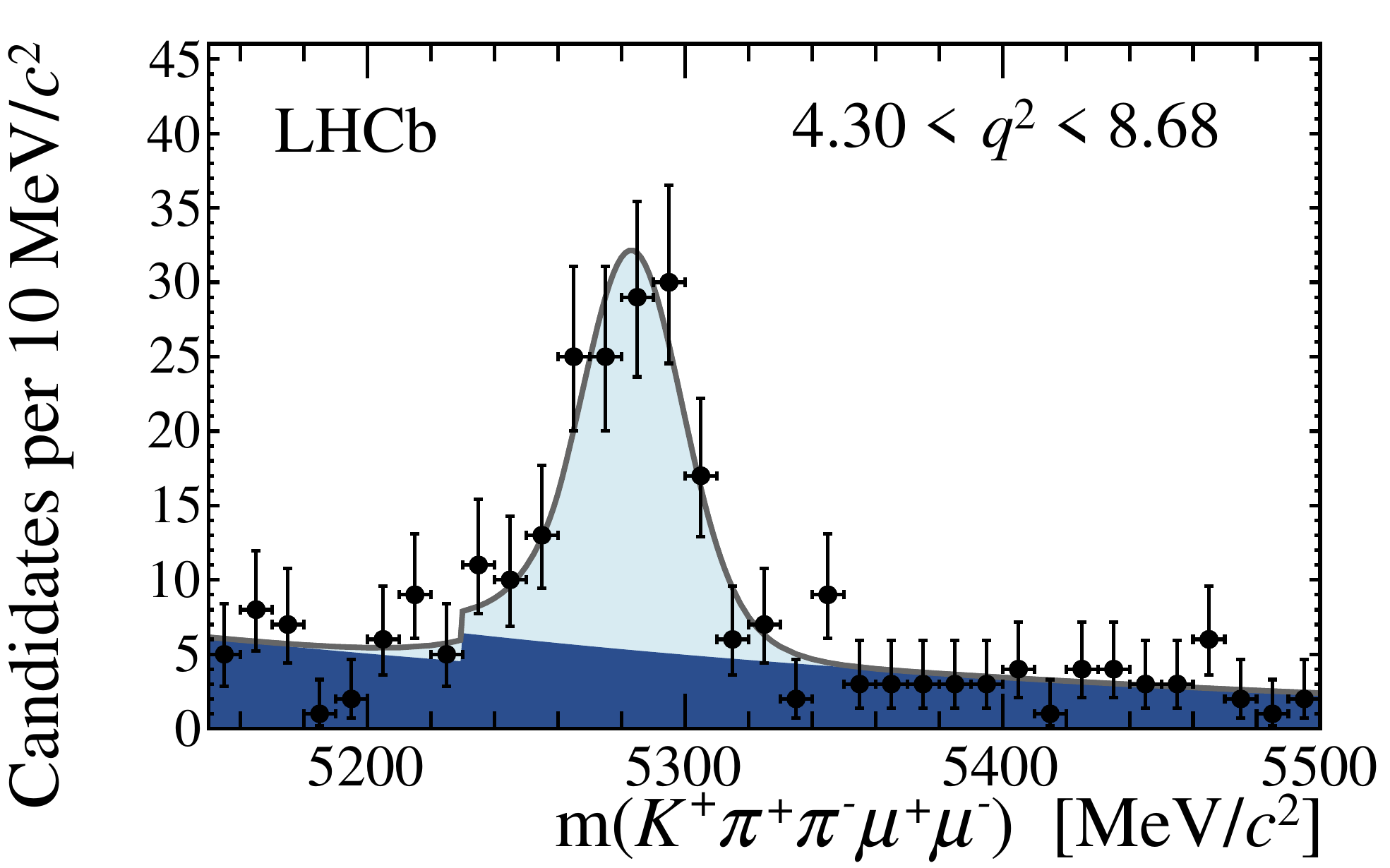}
    \includegraphics[width=0.49\textwidth]{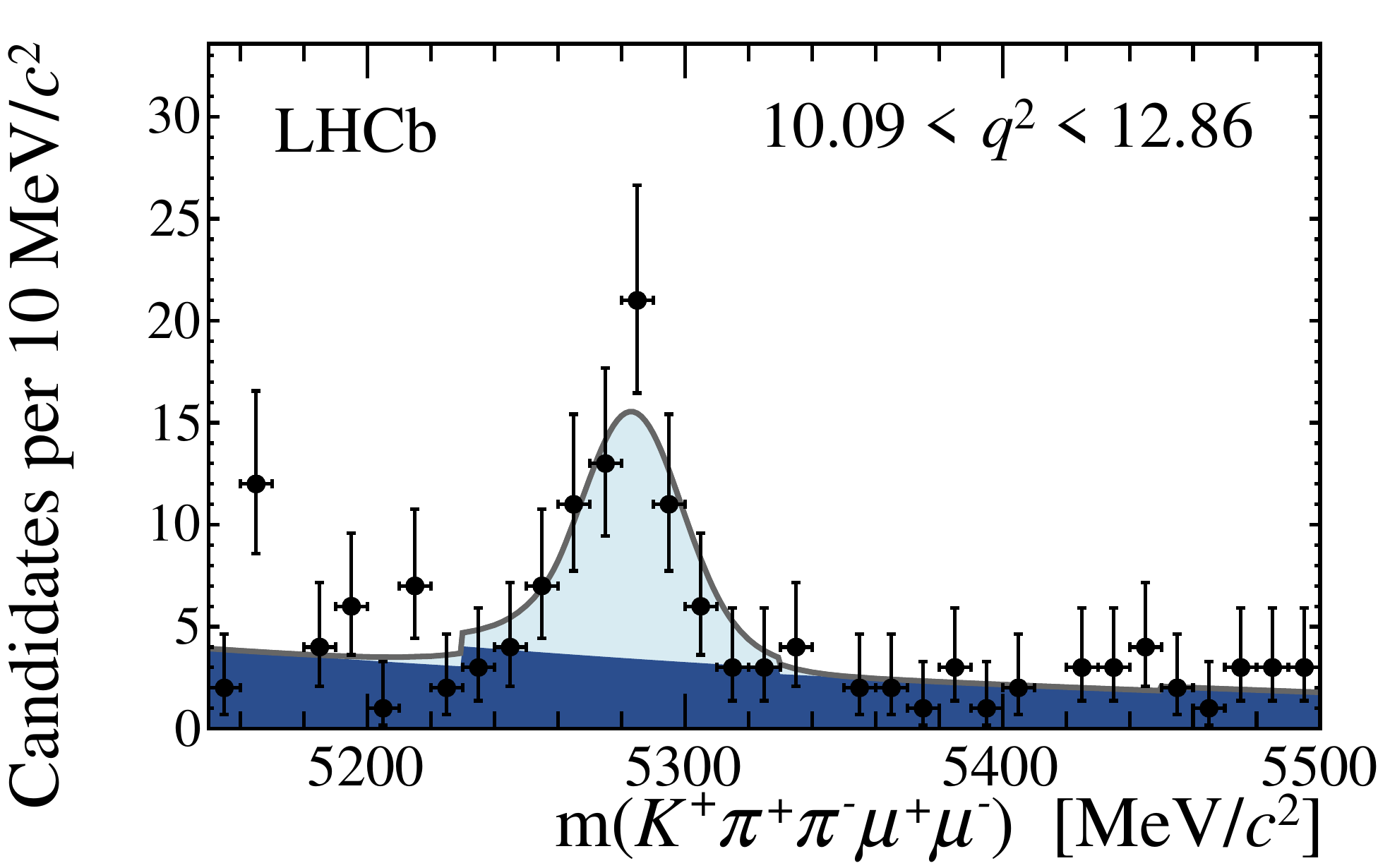}\\
    \includegraphics[width=0.49\textwidth]{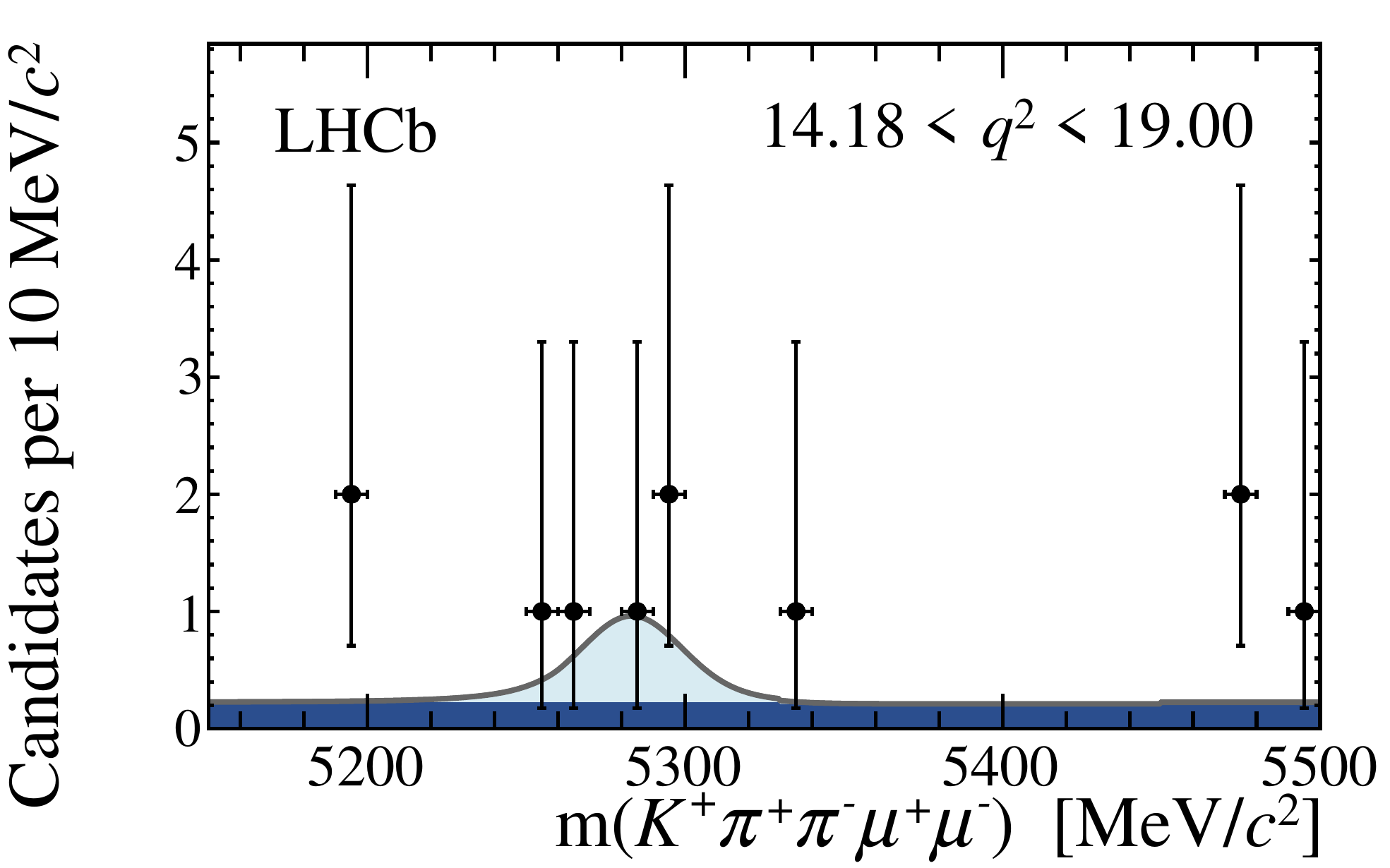}
    \includegraphics[width=0.49\textwidth]{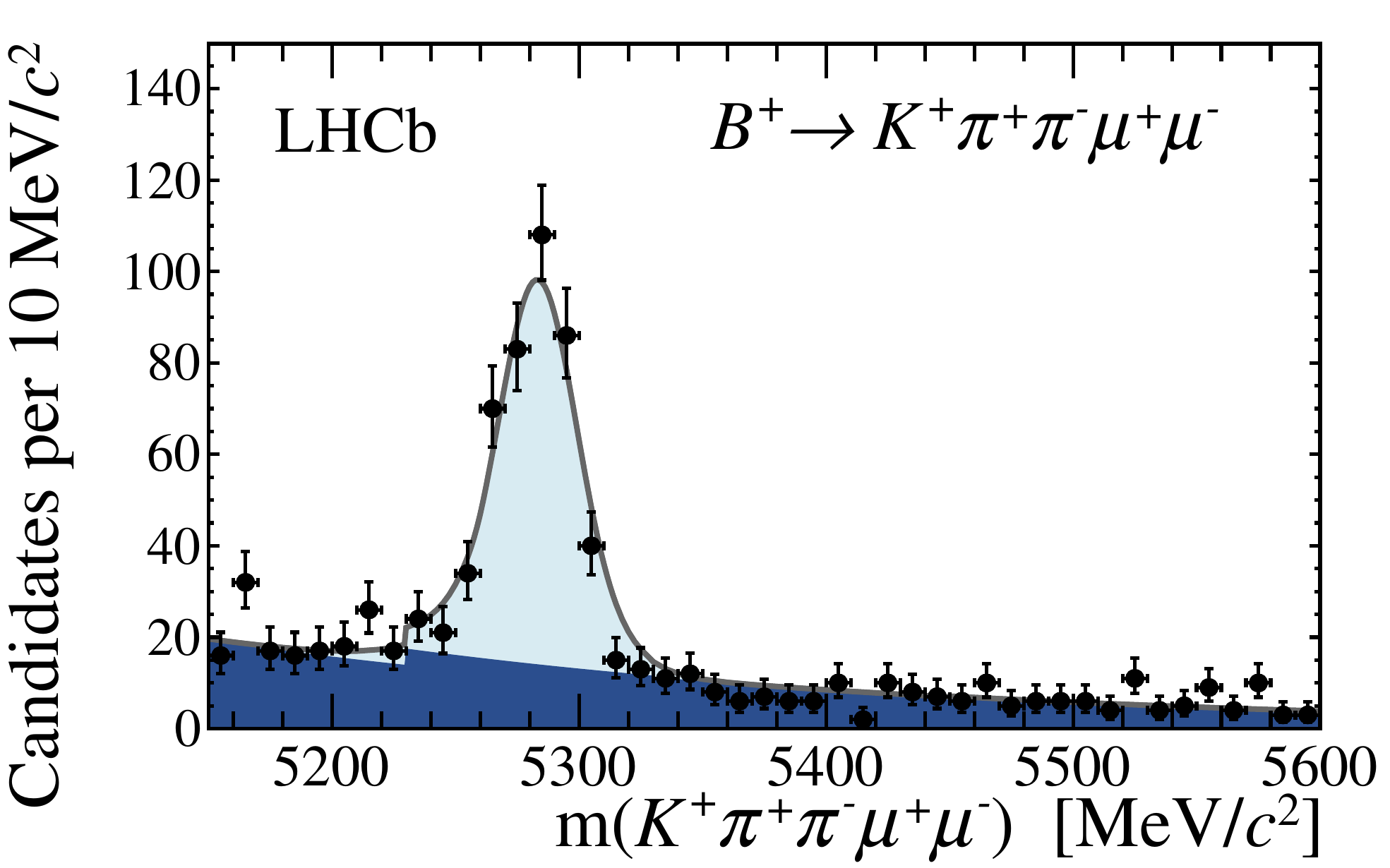}
    \caption{\small Invariant mass of \btokpipimumu candidates
      in bins of $q^2$ with fit projections overlaid.
      The signal component (shaded light blue) is modelled by the sum of two Gaussian functions, each with
      a power-law tail at low mass. The background component (shaded dark blue) is modelled by an exponential function.
      In the $q^2$ ranges $4.30<\qsq<8.68$\gevgevcccc,
      $10.09<\qsq<12.86$\gevgevcccc, and
      $14.18<\qsq<19.00$\gevgevcccc, scaling factors are applied
      to account for the vetoes of the radiative tails of the charmonium
      resonances, resulting in steps in the background mass shape. 
      The lower right plot shows a separate fit to the signal decay integrated over all $q^2$ bins. 
    }
    \label{fig:q2}
  \end{center}
\end{figure}

\begin{figure}
  \begin{center}
  \includegraphics[width=0.49\textwidth]{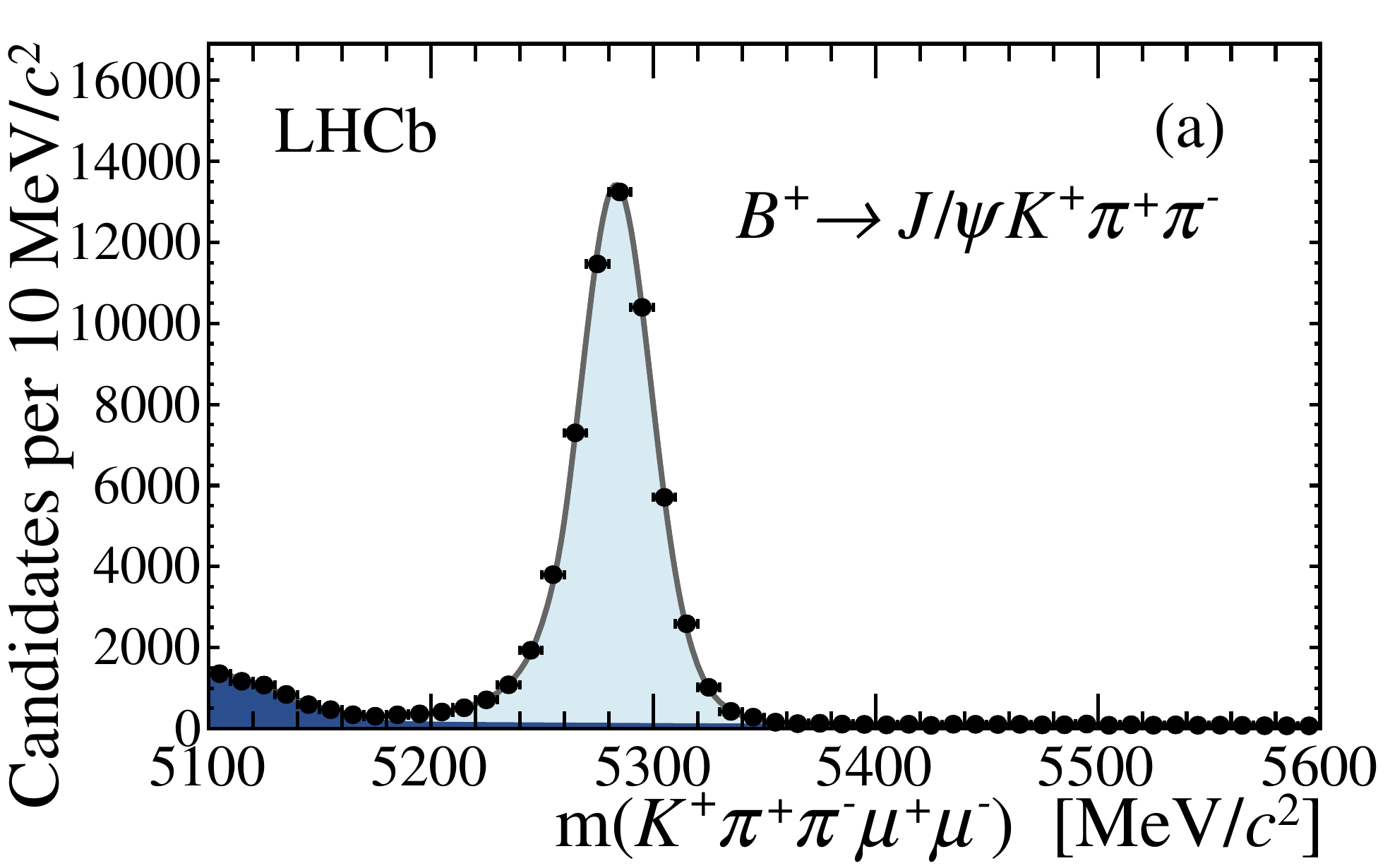}
  \includegraphics[width=0.49\textwidth]{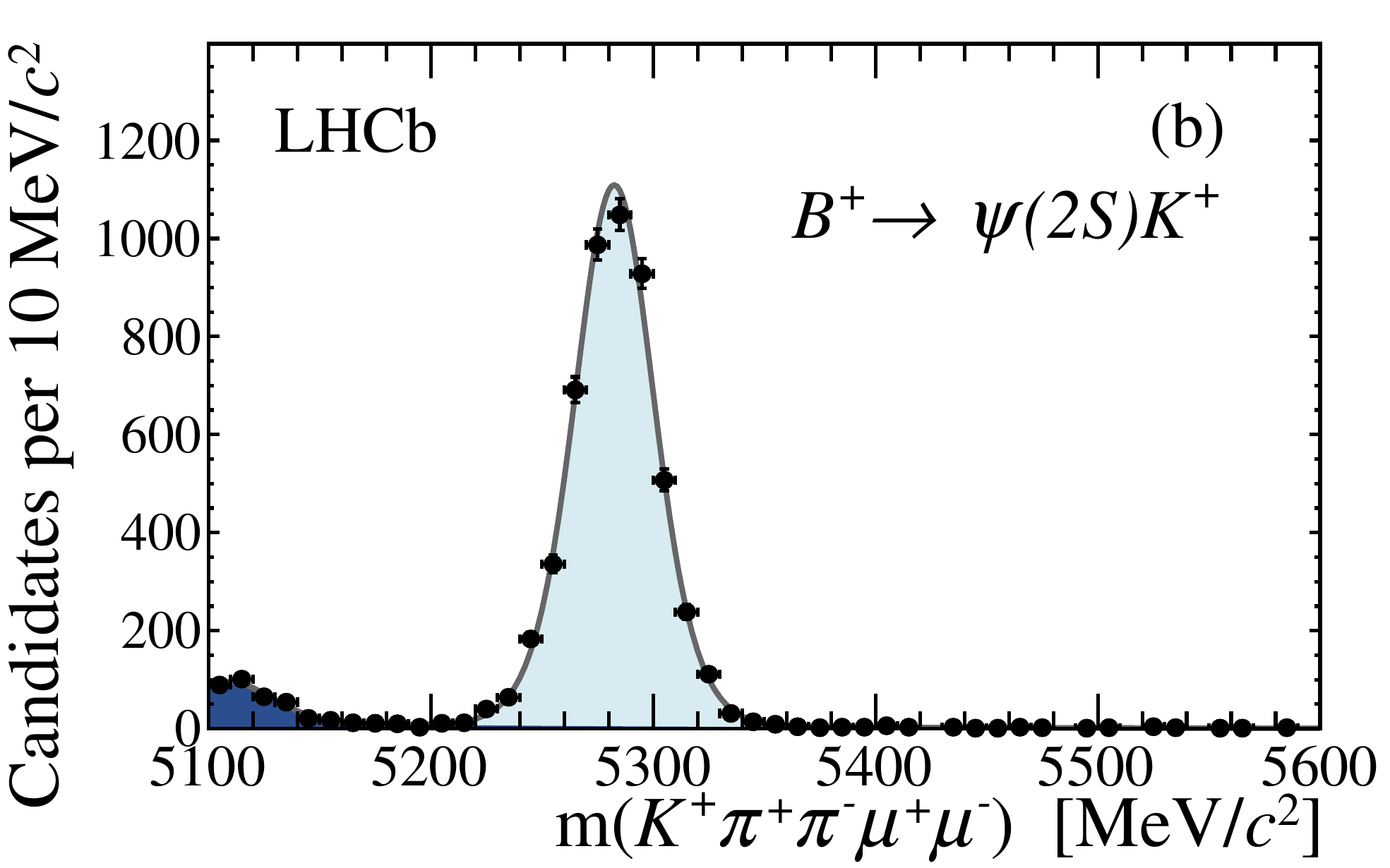}
    \caption{\small Invariant mass distribution of 
(a) the control decay \btokpipijpsi 
and (b) the normalisation mode \btopsitwosk  
with fit projections overlaid. 
    }
    \label{fig:norm}
  \end{center}
\end{figure}

The determination of the differential branching fraction $\dBF(\btokpipimumu)/\dqsq$ is
performed in bins of $q^2$, as given in Table~\ref{tab:diffbf}.
Figure~\ref{fig:q2} shows the invariant mass distribution of $\btokpipimumu$ candidates in each $q^2$ bin studied. 
Signal yields are determined using extended maximum likelihood fits to the
unbinned $\kpipi\mumu$ mass spectra. 
The $m(\kpipi\mumu)$ distribution of the signal component is modelled using the
sum of two Gaussian functions, each with a power-law tail on the low-mass side.
The background component is modelled with an exponential function, where
the reductions in efficiency due to the vetoes of the radiative tails of the charmonium decays are accounted for by using scale factors.
The signal yield integrated over the full $q^2$ range is $N_{K\pi\pi\mu\mu}=367\,_{-23}^{+24}$.
The statistical significance of the signal 
is in excess of 20 standard deviations, 
according to Wilks' theorem~\cite{ref:wilks}. 
Figure~\hyperref[fig:norm]{2a}
shows the fit to the mass distribution of the control channel \btojpsikpipi\ 
that is used to determine the parameters describing the mass distribution of the \btokpipimumu\ signal decay. 
To account for partially reconstructed decays at low masses, a Gaussian function is used in addition to the exponential to describe the background component.
The yield of the control channel is $59\,335\pm343$.
Figure~\hyperref[fig:norm]{2b}
shows the fit for the normalisation channel $\btopsitwosk$. 
To describe the mass shape, 
the same components are used as for the fit of the control decay 
and all mass shape parameters are allowed to vary in the fit. 
The yield of the normalisation channel is $5128\pm67$.

The differential branching fraction $\dBF(\btokpipimumu)/\dqsq$ in 
a $q^2$ bin of width $\Delta q^2$ is
\begin{align}
  \frac{\dBF(\btokpipimumu)}{\dqsq} & = 
\frac{1}{\Delta q^2}\cdot
  \frac{N_\sig}{N_\norm} \cdot
  \frac{\epsilon_\norm}{\epsilon_\sig}\cdot
  \BF\left(\btopsitwosk\right) \nonumber\\
& \hphantom{=}
\cdot\BF\left(\decay{\psitwos}{\jpsi(\to\mumu)\pipi}\right).
\end{align}
Here, $N_\sig$ is the yield of the signal channel in the given $q^2$ bin and $N_\norm$ the yield of the normalisation channel.
The efficiencies for the reconstruction and selection of the signal and normalisation channels are denoted by $\epsilon_\sig$ and $\epsilon_\norm$, respectively.
The efficiency for the signal decay is determined using simulated $\btokonemumu$ 
events generated according to Ref.~\cite{Hatanaka:2008gu}; 
a separate efficiency ratio is calculated for each $q^2$ bin.
The branching fraction for the $\psitwos$ meson to decay to the final state
$\pip\pim\mup\mun$ is
$\BF(\decay{\psitwos}{\jpsi(\to \mup\mun)\pip\pim})=(2.016\pm 0.031)\times 10^{-2}$~\cite{PDG2012}.

{\renewcommand{\arraystretch}{1.2}
\begin{table}
  \begin{center}
    \caption{\small Signal yields for the decay $\btokpipimumu$ 
    and resulting differential branching fractions in bins of \qsq. 
    The first contribution to the uncertainty is statistical, the second systematic, 
where the uncertainty due to the branching fraction of the normalisation channel is included. 
The $q^2$ binning used corresponds to the binning used in previous analyses of $b\to s\mup\mun$ 
decays~\cite{LHCb-PAPER-2013-017,LHCb-PAPER-2013-037,LHCb-PAPER-2013-019}. 
Results are also presented for the $q^2$ range from $1$ to $6\gevgevcccc$, where theory predictions are expected to be most reliable. 
    }
    \begin{tabular}{ccc}\hline
      \qsq bin $[\gevgevcccc]$  & $N_\sig$ & $\tfrac{\dBF}{\dqsq}\;[\e{-8}\pergevgevcccc]$
      \\[0.3ex]\hline\hline
      $[\pz0.10,\pz2.00]$ & $134.1\,^{+12.9}_{-12.3}$     & $7.01\,^{+0.69}_{-0.65} \pm 0.47$ \\
      $[\pz2.00,\pz4.30]$ & $\pz56.5\,^{+\pz9.7}_{-\pz9.1}$ & $2.34\,^{+0.41}_{-0.38} \pm 0.15$ \\
      $[\pz4.30,\pz8.68]$ & $119.9\,^{+14.6}_{-13.7}$     & $2.30\,^{+0.28}_{-0.26} \pm 0.20$ \\
      $[10.09,12.86]$     & $\pz54.0\,^{+10.1}_{-\pz9.4}$   & $1.83\,^{+0.34}_{-0.32} \pm 0.17$ \\
      $[14.18,19.00]$     & $\pzz3.3\,^{+\pz2.8}_{-\pz2.1}$ & $0.10\,^{+0.08}_{-0.06} \pm 0.01$ \\ \hline
      $[\pz1.00,\pz6.00]$ & $144.8\,^{+14.9}_{-14.3}$     & $2.75\,^{+0.29}_{-0.28} \pm 0.16$ \\\hline
    \end{tabular}
    \label{tab:diffbf}
  \end{center}
\end{table}}

\begin{figure}
  \begin{center}
    \includegraphics[width=0.60\textwidth]{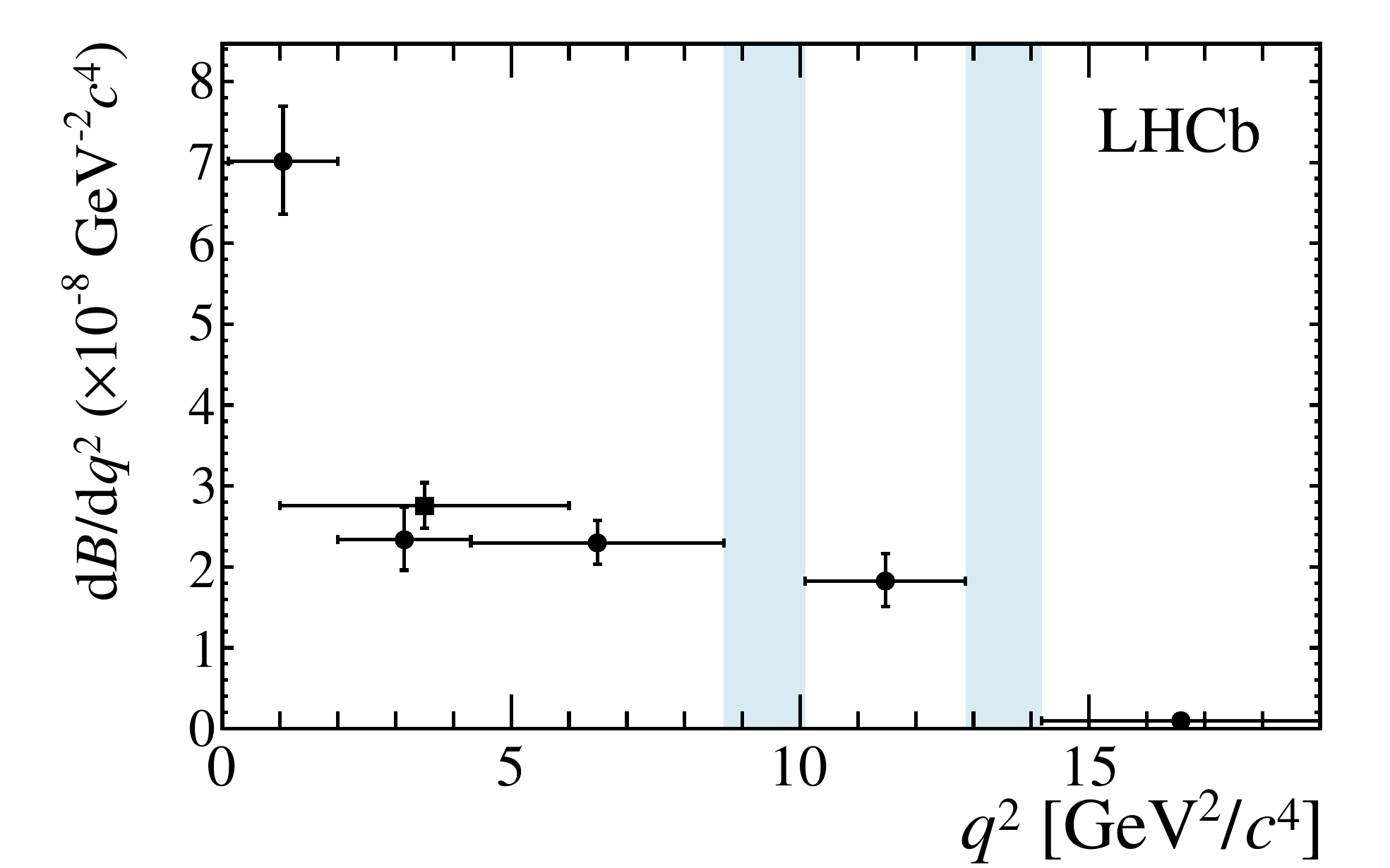}
    \caption{\small Differential branching fraction
      $\dBF(\btokpipimumu)/\dqsq$.
      Errors shown include both statistical and systematic uncertainties.
      Shaded regions indicate 
      the vetoed charmonium resonances.
    }
    \label{fig:diffbf}
  \end{center}
\end{figure}

The resulting differential branching fractions for the decay $\btokpipimumu$ are shown in
Fig.~\ref{fig:diffbf} with numerical values given in Table~\ref{tab:diffbf}.
Summation over all $q^2$ bins yields an integrated branching fraction of
$\left(3.43\,^{+0.23}_{-0.21}\,\mathrm{(stat)}\pm0.15\,\mathrm{(syst)}\pm0.14\,\mathrm{(norm)}\right)\e{-7}$, 
where the uncertainties are statistical, systematic, and due to the uncertainty on the normalisation channel. 
The fraction of signal events removed by the vetoes of the charmonium regions
is determined from simulated $\btokonemumu$ events to be $(21.3\pm 1.5)\%$. 
The uncertainty on this number is determined from a variation of the angle $\thetakone$ and the form-factor parameters within their uncertainties. 
Correcting for the charmonium vetoes yields a total branching fraction of
\begin{align*}
  \BF(\btokpipimumu) &= \left(4.36\,^{+0.29}_{-0.27}\,\mathrm{(stat)}\pm 0.21\,\mathrm{(syst)}\pm0.18\,\mathrm{(norm)}\right) \e{-7}.
\end{align*}
Since the systematic uncertainty due to the normalisation channel is significant, we also report the branching ratio of the signal channel with respect to its normalisation mode, which is determined to be
\begin{align*}
  \frac{\BF(\btokpipimumu)}{\BF(\btopsitwosk)} &= \left(6.95\,^{+0.46}_{-0.43}\,\mathrm{(stat)}\pm0.34\,\mathrm{(syst)}\right)\times 10^{-4}.
\end{align*}

\begin{figure}
  \begin{center}
    \includegraphics[width=0.49\textwidth]{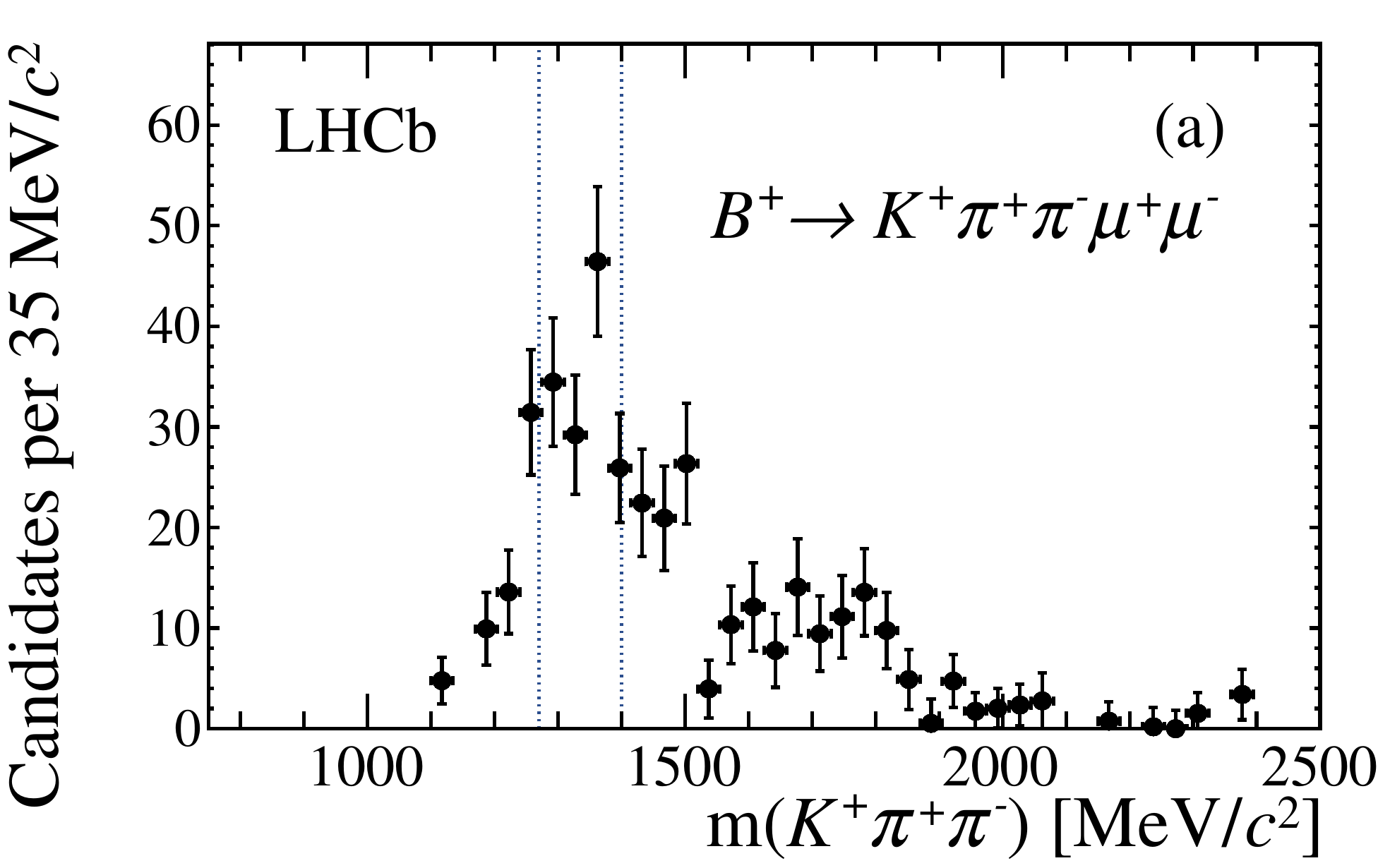}
    \includegraphics[width=0.49\textwidth]{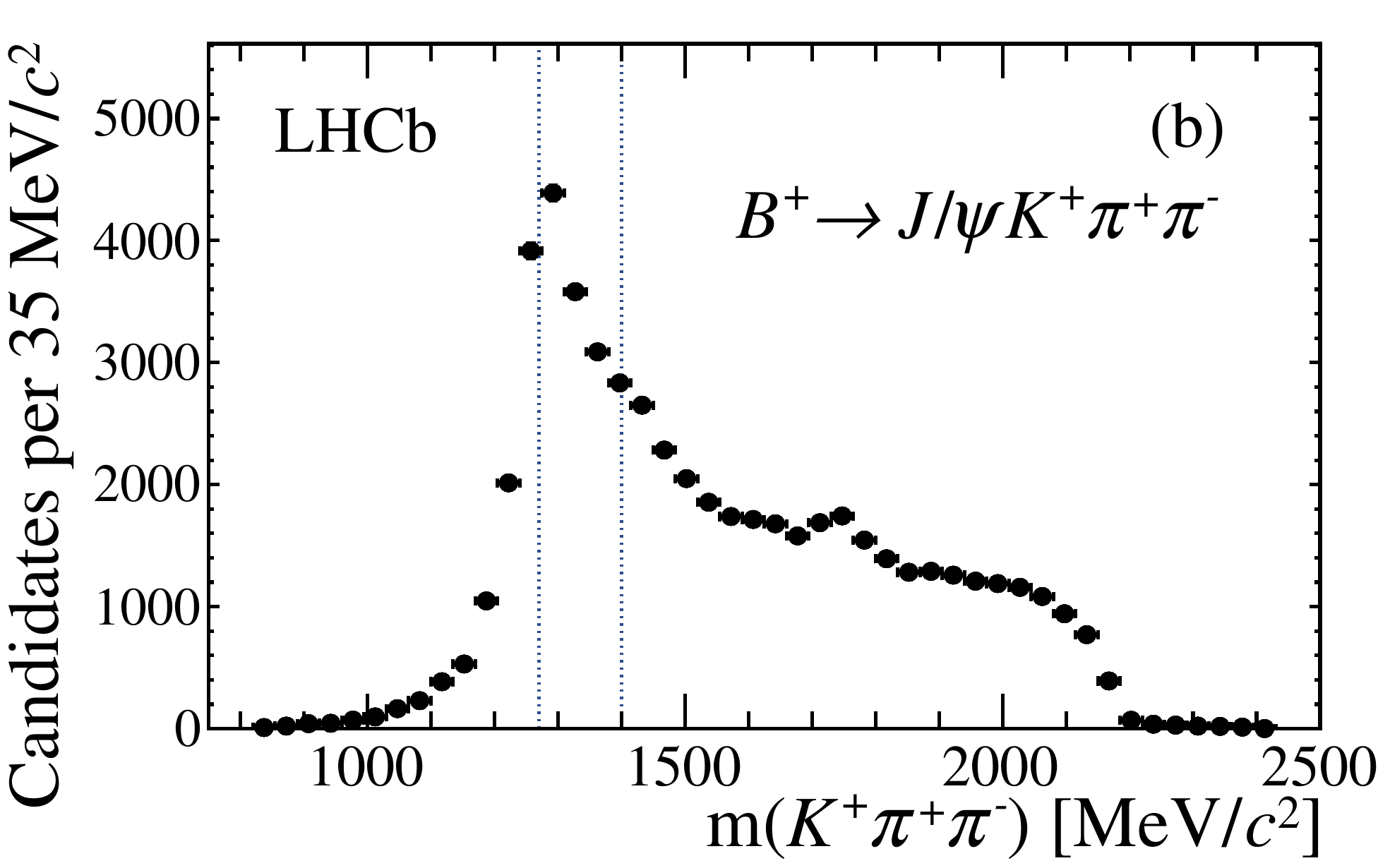}
    \caption{\small
      Background-subtracted $m(\Kp\pip\pim)$ distributions for (a) the signal decay
      $\btokpipimumu$ and (b) the control channel $\btojpsikpipi$.
      The vertical lines indicate the masses of the $K_1(1270)^+$ and $K_1(1400)^+$ resonances.
    }
    \label{fig:kpipi}
  \end{center}
\end{figure}

Due to the low signal yield, no attempt is made to resolve the different contributions to the $\Kp\pip\pim$ system in the $\Kp\pip\pim\mup\mun$ final state.
However, it is possible to obtain the $m(\Kp\pip\pim)$ distribution using the {\it sPlot}~\cite{2005NIMPA.555..356P} technique.
Figure~\ref{fig:kpipi} shows this distribution 
for the signal decay in the full $q^2$ region, as well as for the control decay $\btojpsikpipi$. 
For the signal decay $\btokpipimumu$ the data are consistent with the presence of several broad and overlapping resonances. 

\subsection{Systematic uncertainties}
\label{sec:kpipisyst}
The dominant systematic uncertainty comes from the branching fraction of the normalisation mode $\btopsitwosk$, which is known to a precision of $6\%$.
This uncertainty is fully correlated between the $q^2$ bins and is quoted separately. 

The systematic uncertainty introduced by the choice of signal mass model is estimated by re-evaluating the signal yield using
a single Gaussian function with a power-law tail.
To estimate the uncertainty of the background mass model, a linear mass shape is used instead of the nominal exponential function.
The total systematic uncertainty assigned due to the modelling of the mass distribution is approximately 2\%.

The majority of systematic effects bias the efficiency ratio
$\epsilon_\norm/\epsilon_\sig$, which is determined using simulation.
To account for differences between data and simulation, corrections based on data are applied to simulated events.
The efficiency to identify kaons is corrected by using large $\decay{\Dstarp}{\Dz(\to\Km\pip)\pip}$ control samples. 
Muon identification performance and tracking efficiency are corrected using $\decay{\jpsi}{\mup\mun}$ decays. 
In addition, track multiplicity and vertex fit quality are weighted according to the control channel $\btojpsikpipi$. 
The systematic uncertainties associated with these corrections are evaluated by determining
the branching fraction without the correction and taking the full observed deviation as a systematic uncertainty.
In total, they constitute a systematic uncertainty of around 1\%.
The software trigger is observed to be well described in simulation, but slight discrepancies are observed for the hardware stage.
These are corrected by weighting the simulated samples according to the maximum muon \pt.
The branching fraction is recalculated without these weights, and the observed difference of 1\% is assigned as the systematic uncertainty from the trigger simulation. 

Additional systematic uncertainties stem from the fact that simulated $\btokonemumu$ events, 
modelled according to Ref.~\cite{Hatanaka:2008gu}, are used to determine the efficiency ratio $\epsilon_\norm/\epsilon_\sig$.
To account for contributions other than the $K_1(1270)^+$ to the $\Kp\pip\pim$ system, events are weighted according to the $m(\Kp\pip\pim)$ distribution 
shown in Fig.~\ref{fig:kpipi}. 
This results in a systematic uncertainty of 1--2\%, depending on the \qsq range
considered.
The effect of a potentially different $q^2$ distribution of the signal decay is evaluated by defining the efficiency ratio using $\btokonemumu$ events generated according to a phase-space model.
The observed deviation results in a systematic uncertainty of 1--2\%.

\section[Branching fraction of the decay \btophikmumu]
{Branching fraction of the decay
  \mbox{$\boldsymbol{B^+\!\rightarrow\phi K^+\mu^+\mu^-}$}}
\label{sec:phik}
\begin{figure}
  \begin{center}
    \includegraphics[width=0.49\textwidth]{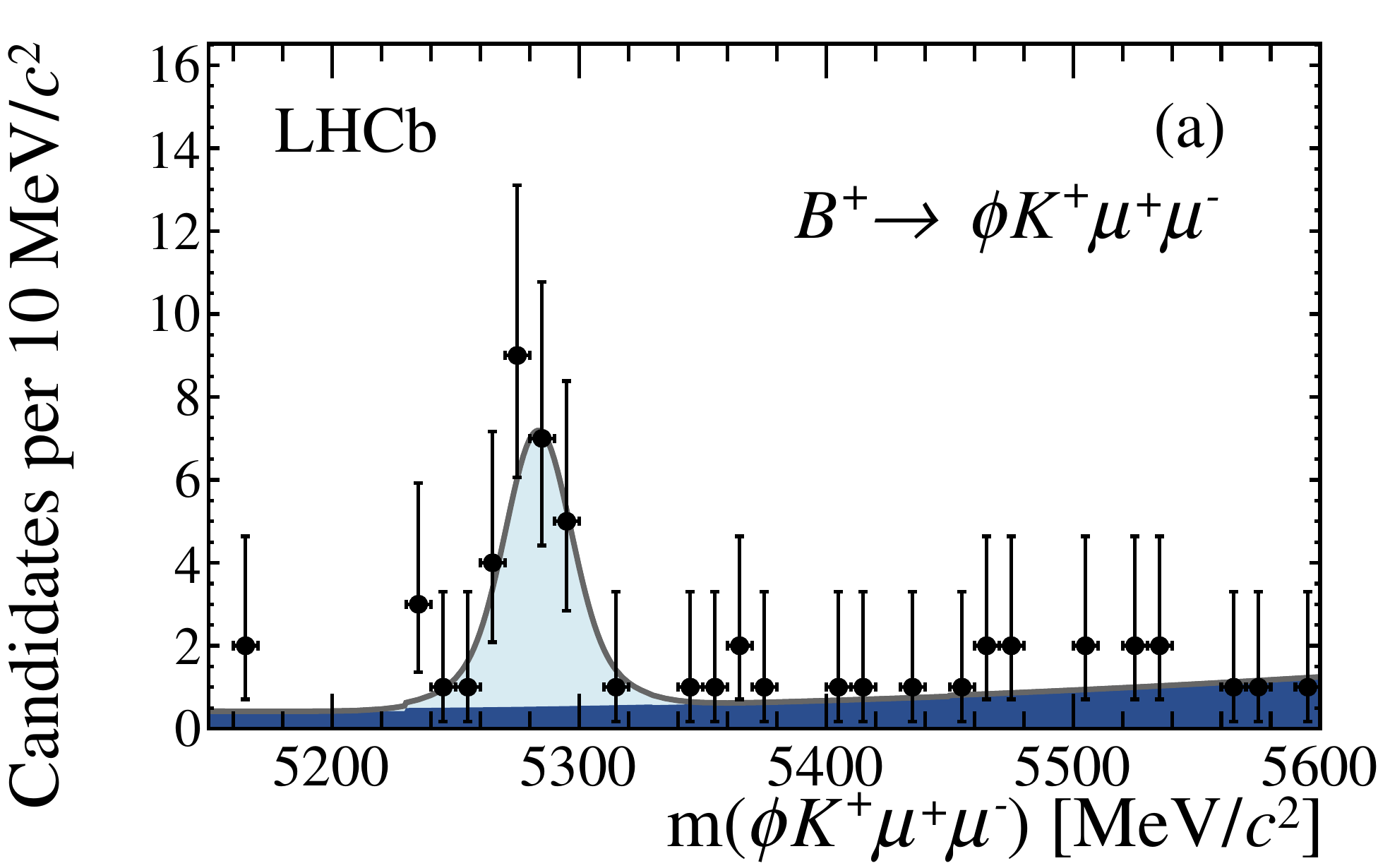}
    \includegraphics[width=0.49\textwidth]{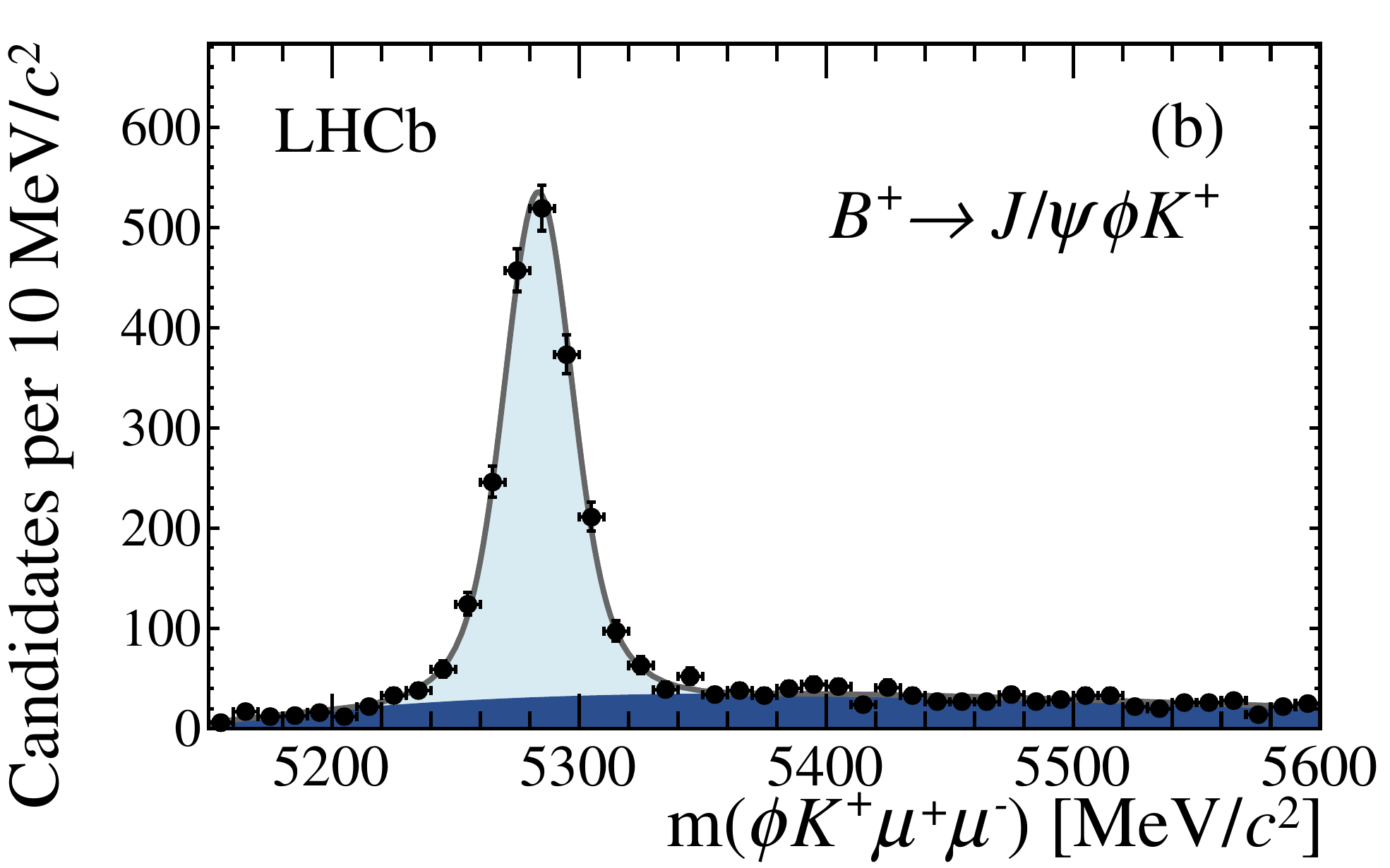}
  \end{center}
  \caption{\small
Invariant $m(\phi\Kp\mumu)$ distributions for 
    (a)~\btophikmumu and (b)~\btophikjpsi\ decays with fit projections overlaid. 
  }
  \label{fig:kphi}
\end{figure}

The signal decay $\btophikmumu$ is expected to be rarer than the decay $\btokpipimumu$ as 
an $s\bar{s}$ quark pair must be created from the vacuum. 
Therefore, only the total branching fraction of this decay mode is determined.
Figure~\hyperref[fig:kphi]{5a} 
shows the $\btophikmumu$ signal candidates after the full selection.
The signal yield is determined to be $N_\sig=25.2\,^{+6.0}_{-5.3}$ using an extended maximum likelihood fit to the unbinned $\phi\Kp\mumu$ mass distribution.
The statistical significance of the signal, calculated using Wilks' theorem, is $6.6\,\sigma$.
The signal component is modelled using the sum of 
two Gaussian functions with a tail described by a power law on the low-mass side.
The background mass shape is modelled using a second-order Chebychev polynomial.
The parameters describing the signal mass shape are fixed to those determined using the normalisation mode $\btojpsiphik$, 
as shown in Fig.~\hyperref[fig:kphi]{5b}. 
The yield of the normalisation mode is $N_\norm=1908\pm63$.

To determine the total branching fraction of the decay $\btophikmumu$, the formula 
\begin{align}
  \BF(\btophikmumu) &= \frac{N^\prime_\sig}{N_\norm} \cdot \BF(\btophikjpsi) \cdot \BF(\decay{\jpsi}{\mumu})
\end{align}
is used.
Here, $N^\prime_\sig$ denotes the signal yield determined in a fit where signal candidates
are weighted by the relative efficiency $\epsilon_\norm/\epsilon_\sig(\qsq)$, according to their $q^2$ value.
This is necessary since the efficiency ratio varies significantly over the full $q^2$ range.
The weights are determined in bins of $q^2$, 
with the same choice of $q^2$ bins as in Table~\ref{tab:diffbf}. 
Using the branching fraction of the normalisation channel, 
the integrated branching fraction is determined to be 
$\left(0.81\,^{+0.18}_{-0.16}\,\mathrm{(stat)}\pm 0.03\,\mathrm{(syst)}\pm0.27\,\mathrm{(norm)}\right)\e{-7}$. 
The fraction of signal events rejected by the charmonium vetoes is $(2\,_{-\pz2}^{+10})\%$. 
This is 
calculated using simulated $\btophikmumu$ events generated according to a phase-space model.  
The uncertainty is estimated by comparison with the model 
given in Ref.~\cite{Hatanaka:2008gu} for the decay $\btokonemumu$ 
and weighting to correct for the large mass of the $\phi\Kp$ system. 
Accounting for the charmonium vetoes 
results in a total branching fraction of
\begin{align*}
  \BF(\btophikmumu) &= \left(0.82 \,^{+0.19}_{-0.17}\,\mathrm{(stat)}\,^{+0.10}_{-0.04}\,\mathrm{(syst)} \pm 0.27\,\mathrm{(norm)}\right)  \times10^{-7}.
\end{align*}
The branching fraction of the signal channel with respect to its normalisation mode is determined to be
\begin{align*}
  \frac{\BF(\btophikmumu)}{\BF(\btojpsiphik)} &= \left(1.58\,^{+0.36}_{-0.32}\,\mathrm{(stat)}\,^{+0.19}_{-0.07}\,\mathrm{(syst)}\right)\times 10^{-3}.
\end{align*}

\subsection{Systematic uncertainties}
The main systematic uncertainty arises from the measurement of the branching fraction of the
normalisation channel, which is known to 33\%~\cite{PDG2012}. 
The systematic uncertainty due to the choice of signal mass model is determined
by using a single Gaussian function with power-law tail on the low-mass side to determine the signal yield. 
For the background mass model, a first-order polynomial, instead of the nominal second-order polynomial, is used.
The total systematic uncertainty from the model used to describe the $m(\phi\Kp\mumu)$ distribution is $3\%$.

The majority of the systematic uncertainties affect the efficiency ratio
$\epsilon_\norm/\epsilon_\sig(q^2)$ and arise from the corrections
based on data that are applied to simulation, as described in Sec.~\ref{sec:kpipisyst}. 
The systematic uncertainties caused by these corrections 
are determined to be $1\%$ in total. 
The limited size of the simulated samples available to calculate the efficiency
ratio introduces an uncertainty of $1.5\%$.
Imperfect modelling of the hardware trigger is corrected for
in the same way as for the measurement of $\BF(\btokpipimumu)$ 
in Sec.~\ref{sec:kpipi} and results in a systematic uncertainty of $1.5\%$.

The efficiency ratio $\epsilon_\norm/\epsilon_\sig(q^2)$ is determined using simulated $\btophikmumu$ 
events generated according to a phase-space model.
The uncertainty due to the $q^2$ distribution in the bins is evaluated by weighting simulated events to reproduce the $q^2$ distribution of $\btokonemumu$ decays. 
This leads to a systematic uncertainty of $1.5\%$.

\section{Conclusions}
\label{sec:conc}
First observations of the rare $b\to s$ FCNC decays \btokpipimumu\ and \btophikmumu\ are presented.
Their branching fractions are measured to be
\begin{align*}
  \mathcal{B}(B^+\rightarrow K^+\pi^+\pi^-\mu^+\mu^-) =&
  \left(4.36\,^{+0.29}_{-0.27}\,\mathrm{(stat)}\pm 0.21\,\mathrm{(syst)}\pm 0.18\,\mathrm{(norm)}
  \right)\times10^{-7},\\
  \mathcal{B}(B^+\rightarrow \phi K^+\mu^+\mu^-) =&
  \left(0.82\,^{+0.19}_{-0.17}\,\mathrm{(stat)}\,^{+0.10}_{-0.04}\,\mathrm{(syst)}\pm 0.27\,\mathrm{(norm)}
  \right)  \times10^{-7},
\end{align*}
where the first uncertainties are statistical, the second systematic and the third due to the uncertainties on the normalisation channels. 
Accounting for the branching fraction 
$\BF(\decay{K_1(1270)^+}{\Kp\pip\pim})= (35.7\pm 3.7)\%$~\cite{PDG2012}, 
the measured branching
fraction for the decay $\btokpipimumu$ is lower than, but compatible with, the SM prediction 
of $\BF(\btokonemumu) = (2.3\,^{+1.3}_{-1.0}\,^{+0.0}_{-0.2})\e{-6}$~\cite{Hatanaka:2008gu}. 
For the decay $\btokpipimumu$, 
the differential branching fraction
$\dBF(\btokpipimumu)/\dqsq$ is also determined.

% Do not include this in analysis note and conference reports
\section*{Acknowledgements}
%
%The text below are the acknowledgements as approved by the collaboration
%board. Extending the acknowledgements to include individuals from outside the
%collaboration who have contributed to the analysis should be approved by the
%EB and, if possible, be included in the draft of first circulation.
%
%
\noindent We express our gratitude to our colleagues in the CERN
accelerator departments for the excellent performance of the LHC. We
thank the technical and administrative staff at the LHCb
institutes. We acknowledge support from CERN and from the national
agencies: CAPES, CNPq, FAPERJ and FINEP (Brazil); NSFC (China);
CNRS/IN2P3 (France); BMBF, DFG, HGF and MPG (Germany); SFI (Ireland); INFN (Italy); 
FOM and NWO (The Netherlands); MNiSW and NCN (Poland); MEN/IFA (Romania); 
MinES and FANO (Russia); MinECo (Spain); SNSF and SER (Switzerland); 
NASU (Ukraine); STFC (United Kingdom); NSF (USA).
The Tier1 computing centres are supported by IN2P3 (France), KIT and BMBF 
(Germany), INFN (Italy), NWO and SURF (The Netherlands), PIC (Spain), GridPP 
(United Kingdom).
We are indebted to the communities behind the multiple open 
source software packages on which we depend. We are also thankful for the 
computing resources and the access to software R\&D tools provided by Yandex LLC (Russia).
Individual groups or members have received support from 
EPLANET, Marie Sk\l{}odowska-Curie Actions and ERC (European Union), 
Conseil g\'{e}n\'{e}ral de Haute-Savoie, Labex ENIGMASS and OCEVU, 
R\'{e}gion Auvergne (France), RFBR (Russia), XuntaGal and GENCAT (Spain), Royal Society and Royal
Commission for the Exhibition of 1851 (United Kingdom).

% This should be taken out in the final paper
%\input{supplementary-app}

\addcontentsline{toc}{section}{References}
\setboolean{inbibliography}{true}
\bibliographystyle{LHCb}
\bibliography{main,LHCb-PAPER,LHCb-CONF,LHCb-DP,LHCb-TDR}

\newpage
%%%%%%%%%%%%%%%%%%%%%%%%%%%%%%%%%%%%%%%%%%
\centerline{\large\bf LHCb collaboration}
\begin{flushleft}
\small
R.~Aaij$^{41}$, 
B.~Adeva$^{37}$, 
M.~Adinolfi$^{46}$, 
A.~Affolder$^{52}$, 
Z.~Ajaltouni$^{5}$, 
S.~Akar$^{6}$, 
J.~Albrecht$^{9}$, 
F.~Alessio$^{38}$, 
M.~Alexander$^{51}$, 
S.~Ali$^{41}$
% ...

{\footnotesize \it
% ...
$ ^{5}$Clermont Universit\'{e}, Universit\'{e} Blaise Pascal, CNRS/IN2P3, LPC, Clermont-Ferrand, France\\
$ ^{6}$CPPM, Aix-Marseille Universit\'{e}, CNRS/IN2P3, Marseille, France\\
% ...
$ ^{9}$Fakult\"{a}t Physik, Technische Universit\"{a}t Dortmund, Dortmund, Germany\\
% ...
$ ^{37}$Universidad de Santiago de Compostela, Santiago de Compostela, Spain\\
$ ^{38}$European Organization for Nuclear Research (CERN), Geneva, Switzerland\\
% ...
$ ^{41}$Nikhef National Institute for Subatomic Physics, Amsterdam, The Netherlands\\
% ...
$ ^{46}$H.H. Wills Physics Laboratory, University of Bristol, Bristol, United Kingdom\\
% ...
$ ^{51}$School of Physics and Astronomy, University of Glasgow, Glasgow, United Kingdom\\
$ ^{52}$Oliver Lodge Laboratory, University of Liverpool, Liverpool, United Kingdom\\
% ...
}
\end{flushleft}
%%%%%%%%%%%%%%%%%%%%%%%%%%%%%%%%%%%%%%%%%%

\end{document}